\documentclass{PoS}

\title{Cosmic-ray diffusive reacceleration: a critical look}

\ShortTitle{Cosmic-ray diffusive reacceleration: a critical look}

\author{\speaker{Luke O'C. Drury}\\
        Dublin Institute for Advanced Studies, School of Cosmic Physics,\\
        31 Fitzwilliam Place, Dublin 2, Ireland\\
        E-mail: \email{ld@cp.dias.ie}}

\author{{Andrew W. Strong}\\
Max-Planck-Institut f\"ur Extraterrestrische Physik, Garching, Germany\\
        E-mail: \email{aws@mpe.mpg.de}}

\abstract{Cosmic-ray scattering on magnetic turbulence leads to spatial diffusive propagation; if the scattering medium is moving, this will inevitably also cause changes in the momentum of the particles, so-called diffusive reacceleration. This can be described as diffusion in momentum space. Diffusive reacceleration has often been invoked to explain the peak observed in secondary-to-primary ratios at a few GeV, in particular Boron-to-Carbon. This avoids the necessity to postulate an ad-hoc break in the spatial diffusive coefficient, and has become almost a standard in modelling cosmic-ray spectra. However, at the levels invoked, the process implies a significant input of energy from the interstellar medium into cosmic rays, so that in such models interstellar space competes with the usual accelerators like supernova remnants. The questions arise: is reacceleration really occurring at the high level required to explain secondary-to-primary ratios? and are the energy requirements physically plausible? We address this issue using both analytical and numerical models of cosmic-ray propagation.}

\FullConference{The 34th International Cosmic Ray Conference,\\
		30 July- 6 August, 2015\\
		The Hague, The Netherlands}

\newcommand\Galprop{{\sc Galprop}}

\usepackage{aas_macros}

\begin{document}

\section{Introduction}

In most discussions of cosmic ray propagation it is assumed that the cosmic rays are produced in discrete sources, generally taken to be supernova remnants, although there is increasing interest in and evidence for possible additional classes of sources, see e.g. the presentations in the recent conference \cite{2014NuPhS.256....1T}, and then diffuse through the interstellar medium and an extended cosmic ray halo before escaping from the Galaxy.  This diffusion is not just in space, but also in momentum, if, as is inevitably the case, the magnetic fields that scatter the particles are not purely static but are in random motion with characteristic velocities of order the Alfv\'en speed.  This leads to a certain amount of second order Fermi acceleration during propagation, an effect which for historical reasons is usually referred to as ``reacceleration''.  There has been a long and inconclusive debate as to the importance of reacceleration, but there is no doubt that the idea is attractive for a variety of reasons, and physically the effect must occur at some level. 

One of the strongest arguments in favour of reacceleration is that it allows a rather natural fit to the low energy Boron to Carbon data.  This was persuasively argued by \cite{1995ApJ...441..209H} who showed that with reacceleration the B/C data could be well fit with a single power-law dependence of the escape path-length as a function of rigidity, whereas in a leaky-box model one had to suppose an unnatural decrease in the path-length at low energies.  The fact that the inferred power-law dependence of the path-length also agreed with that predicted for Kolmogorov turbulence was seen as another positive feature of the model (although whether interstellar magnetic turbulence should be described by a Kolmogorov spectrum is another question). 

This reacceleration by interstellar turbulence is a potentially significant contribution to the total Galactic cosmic ray source power (and of course also a potentially significant damping term for interstellar turbulence).   This was qualitatively discussed in \cite{2014MNRAS.442.3010T} on the basis of analytic estimates.  In this paper we seek to make quantitative estimates of the reacceleration power using the well-established cosmic ray propagation code \Galprop\ as first described in \cite{1998ApJ...509..212S}.  \Galprop\ has become the de facto reference model for cosmic ray propagation studies in our Galaxy and is thus an appropriate choice, but any other propagation model incorporating reacceleration should give very similar results. The power transferred to the cosmic rays comes of course at the expense of the general turbulence in the interstellar medium so we also evaluate a damping process for interstellar turbulence.
Finally we note that the whole question of propagation models has acquired new relevance in the context of recent searches for possible signals of dark matter in cosmic rays, see e.g. \cite{2015arXiv150604145K} and references therein.

\section{Reacceleration power}

This section briefly reviews and expands on the main results of \cite{2014MNRAS.442.3010T}.  The basic equation underpinning all diffusion models of cosmic ray transport, going back at least to the classic monograph of  \cite{1964ocr..book.....G} is
\begin{equation}
{\partial f\over\partial t} = Q + 
\nabla\left(D_{xx}\nabla f\right) +  {1\over 4\pi p^2} {\partial\over\partial p}\left(4\pi p^2 D_{pp} {\partial f\over\partial p}\right) +....
\end{equation}
Here $f(x,p,t)$ is the isotropic part of the phase space density of a given species as a function of position $x$, scalar momentum $p$ and time $t$ and $Q(x, p, t)$ is a source term representing the initial production and acceleration of cosmic rays.  The spatial and momentum space diffusion coefficients are $D_{xx}$ and $D_{pp}$ respectively and will in general be functions of position, momentum, and time.  The ellipsis denotes the many other terms that should be included, for example those describing ionisation energy loss and nuclear interactions, but which are irrelevant for this discussion.

On heuristic grounds \cite{2014MNRAS.442.3010T} the two diffusion coefficients are related by
\begin{equation}
D_{xx}D_{pp} \approx {1\over 9} p^2 V_A^2.
\end{equation}
Detailed calculations \cite{1966SvPhU...9..370T,1975MNRAS.172..557S},
see also equations 9.38 and 9.39 in \cite{1990acr..book.....B}, show that
\begin{equation}
D_{xx}D_{pp} =
p^2 V_A^2 
\left<1-\mu^2 \over \nu_+ +\nu_-\right>
\left<(1-\mu^2)\nu_+\nu_-\over\nu_+ + \nu_-\right>
\end{equation}
where the angle brackets denote an average over an isotropic pitch angle distribution $\mu$ being the pitch angle cosine,
\begin{equation}
\left<\phantom{1\over 2}\right> = \int_{-1}^{+1} {d\mu\over 2}
\end{equation}
and $\nu_\pm$ is the rate of pitch-angle scattering off forward and backward propagating Alfv\'en waves.  
If the scattering rates are taken to be constant and equal we recover the naive heuristic result, but in general the scattering rates should be functions of the pitch angle and related to the wave spectrum through the gyro-resonant condition.  This introduces a weak dependence on the shape of the wave spectrum, but not on its amplitude. Specifically, if we assume a power-law wave spectrum with $W(k)\propto k^{-a}$ where $\int W(k)\,dk$ is the total energy content of the waves and $k$ is the wave-number ($a=5/3$ being a Kolmogorov spectrum for example).  
\begin{equation}
\nu_\pm \approx {v\over r_{\rm g}} {k_{\rm res} W_\pm (k_{\rm res})\over B_0^2/2\mu_0}
\end{equation}
where $r_{\rm g} = p/eB_0$ is the particle gyroradius, $B_0$ the background magnetic field,  and the resonant condition is
\begin{equation}
\mu r_{\rm g} k_{\rm res} \approx 1.
\end{equation}

Considered as a function of pitch angle the scattering rates thus scale as
\begin{equation}
\nu_\pm \propto \mu^{a-1}
\end{equation}
so that (assuming $\nu_+= \nu_-$)
\begin{eqnarray}
&&\left<1-\mu^2 \over \nu_+ +\nu_-\right>
\left<(1-\mu^2)\nu_+\nu_-\over\nu_+ + \nu_-\right>\nonumber\\
&=&
{1\over 4} \int_0^1 (1-\mu^2)\mu^{1-a}d\mu \int_0^1(1-\mu^2)\mu^{a-1}d\mu\nonumber\\
&=&
{1\over 4} \left({1\over 2-a} - {1\over 4-a}\right)
\left({1\over a} - {1\over 2+a}\right), \quad 0<a<2, \nonumber\\
&=& {1\over a(4-a)(4-a^2)}.
\end{eqnarray}
Thus more generally
\begin{equation}
D_{xx}D_{pp} =
p^2 V_A^2 {1\over a(4-a)(4-a^2)}
\end{equation}
which agrees with the simple heuristic estimate for $a=1$.  

However if we hold $\mu$ fixed and look at the scattering rate as a function of particle momentum or gyro-radius, then
\begin{equation}
\nu_\pm \propto v r_{\rm g}^{-2} W(k_{\rm res}) \propto v r_{\rm g}^{(a-2)}
\end{equation}
and thus
\begin{equation}
D_{xx} \propto v^2 \left<(1-\mu^2){1\over \nu_+ + \nu_-}\right> \propto v \left(p\over e\right)^{(2-a)}
\end{equation}
which is the commonly used form, $D\propto vR^{\delta}$ or velocity times a power-law in particle rigidity (momentum per charge, which is directly proportional to the gyro-radius for fixed magnetic field).  The power-law index of the rigidity dependence of the spatial diffusion coefficient, $\delta$, is thus related to that of the wave spectrum, $a$, by
\begin{equation}
\delta = 2 - a, \qquad a = 2 -\delta
\end{equation}
and, as is easily verified,
\begin{equation}
{1\over a(4-a)(4-a^2)} = {1\over \delta(4-\delta)(4-\delta^2)}.
\end{equation}
In fact the denominator can be written
\begin{equation}
a(4-a)(4-a^2) = 9 - 10(a-1)^2 + (a-1)^4
\end{equation}
which shows that it is symmetric about and has a local maximum at $a=1$.
Thus within this simple quasilinear theory of particle scattering we have
\begin{equation}
D_{xx}D_{pp} = p^2 V_A^2 {1\over \delta(4-\delta)(4-\delta^2)}\ge {1\over 9} p^2 V_A^2 
\end{equation}
which is essentially the form used by \Galprop, see equation (1) in \cite{1998ApJ...509..212S} derived from \cite{1994ApJ...431..705S}, if their factor $4/3w$ is set to unity (the inclusion of $w$ appears formally incorrect, but as it is treated as a constant of order unity it makes no difference to the results).  It is interesting to note that for equal intensities of forward and backward propagating waves the naive scattering estimate corresponds to the minimum possible level of reacceleration, but of course lower values can be obtained with anisotropic wave fields.

For convenience we will from now on write
\begin{equation}
D_{pp} = \vartheta p^2 V_A^2 {1\over D_{xx}}.
\end{equation}
For an isotropic Kolmogorov spectrum with $\delta= 1/3$ we have $\vartheta = 81/385 \approx 0.21$ and the largest it can be is 
for $\delta = 1$ where we have $\vartheta = 1/9 \approx 0.11$; the results of \cite{1998ApJ...509..212S} can be obtained by setting\begin{equation}
\vartheta = {4\over 3 \delta(4-\delta^2)(4-\delta) w}.
\end{equation}

The local reacceleration power density is shown in  \cite{2014MNRAS.442.3010T}, assuming very reasonable regularity conditions on the particle distribution function, to be given by the integral over the particle spectrum,
\begin{equation}
P_R= \int_0^\infty 4\pi p^2 f {1\over p^2} {\partial\over\partial p} \left(p^2 D_{pp} v\right) \, dp
\end{equation}
which can be written, if $D_{pp}$ is expressed in terms of $D_{xx}$ using the above relation, as
\begin{equation}
P_R =  \int_0^\infty 
4\pi p^2 f \left (\vartheta V_A^2 p v\over D_{xx}\right) \left[ 4 + {\partial\ln(v/D_{xx})\over\partial\ln p}\right]\, dp.
\end{equation}
If we parametrise the spatial diffusion as
\begin{equation}
D_{xx}=D_0 \left(v\over c\right) \left(p\over m c\right)^\delta
\end{equation}
where $m$ is the particle mass and $c$ the speed of light
this can be written in the useful and rather transparent form
\begin{equation}
P_R = \vartheta (4-\delta) {V_A^2\over D_0} m c^2 \int 4\pi p^2 f \left(p\over mc\right)^{1-\delta} dp
\end{equation}
which gives the local reacceleration power density in terms of a simple integral over the spectrum.  Integrating this over the entire Galaxy then gives the total power transferred to the cosmic ray population from Alfv\'enic turbulence.

\subsection{Analytical estimate}

The obvious first step is to consider a power-law distribution function of the form
\begin{equation}
f(p) = f_0 \left(p\over mc\right)^{-\gamma}, \qquad p>p_{\rm min}
\end{equation}
above some minimum momentum $p_{\rm min}$
where observationally $\gamma\approx 4.8$ (the distribution function in momentum is steeper by 2 than the differential energy spectrum, so this corresponds to $N(E)\propto E^{-2.8}$ for relativistic particles). However as is easily seen this gives an integral which is dominated by the low-energy cut-off.
\begin{eqnarray}
P_R &=& \vartheta (4-\delta) {V_A^2\over D_0} m c^2 \int_{p_{\rm min}}^\infty 4\pi p^2 f_0 \left(p\over mc\right)^{1-\delta-\gamma} dp \nonumber\\
&=& \vartheta (4-\delta) {V_A^2\over D_0} m c^2  \,(m c)^3 f_0 \, {1\over \delta+\gamma - 4} \left(p_{\rm min}\over mc\right)^{4-\delta-\gamma}
\end{eqnarray}
(note that $\delta+\gamma >4$) and diverges as $p_{\rm min} \to 0$

%
%
%
%
%
The total reacceleration power is thus dominated by the poorly-know low-energy part of the cosmic ray spectrum, as noted in \cite{2014MNRAS.442.3010T}, but we can at least estimate the contribution from particles above about $1\,\rm GeV/n$ where we have reliable data.

The differential energy flux of cosmic rays above $1\,\rm GeV$ is given by the Particle Data Group booklet as approximately
\begin{equation}
I(E) = 1.8\times 10^{4} \left(E\over 1\,\rm GeV\right)^{-2.7}\,\rm m^{-2}\,s^{-1}\,sr^{-1}\,(GeV)^{-1}
\end{equation}
corresponding to (for simplicity, and as this is a rough estimate, we consider only protons and take the proton rest mass to be $1\,\rm GeV$) a proton distribution function of
\begin{equation}
p^2 f(p) = {I\over c} = 6\times 10^{-5} \left(p\over mc\right)^{-2.7}\,\rm m^{-3}\,sr^{-1}\,(GeV)^{-1}.
\end{equation}
This implies an energy density in relativistic cosmic ray protons of
\begin{eqnarray}
\int 4\pi p^2 f(p)\,pc\, dp &\approx& 6\times 10^{-5} 4\pi mc^2 \int_1^\infty \left(p\over mc\right)^{-1.7} {dp\over 1\,\rm GeV}
\nonumber\\
&\approx& 1.0 \,\rm eV\, cm^{-3}
\end{eqnarray}
and a reacceleration power density for relativistic cosmic rays of
\begin{eqnarray}
P_R &=& \vartheta (4-\delta) {V_A^2\over D_0} m c^2 \int 4\pi p^2 f \left(p\over mc\right)^{1-\delta} dp\nonumber\\
&\approx& \vartheta (4-\delta) {V_A^2\over D_0}{1\over 0.7+\delta}\,{0.75\,\rm eV\,cm^{-3}}\nonumber\\
&=& \vartheta {4-\delta \over 0.7 + \delta} \left(V_A\over 30 \,\rm km\, s^{-1}\right)^2
\left(D_0\over 10^{28}\rm cm^2s^{-1}\right)^{-1} 7\times 10^{-16} \,\rm eV\,cm^{-3}\,s^{-1}
\end{eqnarray}
with more from the low-energy part of the spectrum. 
For a typical CR residence time in the Galaxy of at least $10^7$ yr, reacceleration thus adds at least 20\% to the energy content. To proceed further we need to turn to numerical methods.

%
%
%

\subsection{Estimate using  GALPROP}
We use the numerical CR-propagation package GALPROP\footnote{Current version available at https://sourceforge.net/projects/galprop},
using  parameters which reproduce B/C with reacceleration, and also with the same parameters but without the reacceleration term.
The total CR proton energy content of the Galaxy is computed by integration over momentum and volume. 
The luminosity is computed assuming a nominal residence time in the Galaxy.
We base the calculation on the model z04LMS, which has reacceleration and a CR halo height of 4 kpc \cite{2010ApJ...722L..58S}.
The effective CR residence time is here set to $3\times10^7$ yr to reproduce the  total proton luminosity of  $8\times 10^{40}$   erg   s$^{-1}$  \cite{2010ApJ...722L..58S},
 but the relative values are independent of this assumption.


For  $V_A = 30 \,\rm km \,  s^{-1}$   the proton luminosity is            $7.8 \times 10^{40}\,\rm erg\,  s^{-1}$ 
and for  $V_A$= 40 km    s$^{-1}$ ,              $8.7  \times 10^{40}\,\rm erg\,  s^{-1}$. 
Without reacceleration, the   proton luminosity  is $6 \times 10^{40}\rm\, erg\,  s^{-1}$. 
These are typical parameter values which reproduce B/C in reacceleration models (see also \cite{2011ApJ...729..106T}), so 30-45\% of energy input is from  reacceleration. 
This is consistent with the analytical estimate above.

\section{Conclusions}

Our estimates of the reacceleration power are necessarily approximate, but suffice to demonstrate that, for parameters chosen to reproduce B/C, as is often done,
the energy input via reacceleration  is comparable to the total energy, which is usually considered to arise in SNR. 
SNR are then not ``{\it the} sources of CR'' (in the sense of the sole source of their energy) in such models (although of course they can still contribute indirectly as being among the many agents driving the bulk turbulence of the interstellar medium \cite{2004ARA&A..42..211E})!  The questions of whether this is physically plausible, and what experimental tests could be made to confirm reacceleration or otherwise, are left to a future paper.  We note however that the main dissipation channel for interstellar turbulence is almost certainly the non-linear cascading to small scales (indeed this idea is implicit in the use of Kolmogorov or Kraichnan spectra) and thus the total power requirement for maintaining the interstellar turbulence could be many times the energy going into diffusive reacceleration of cosmic rays.  It is possible that the cascading simply transfers the energy to the relevant scales for reacceleration and that indeed cosmic ray reacceleration is the dominant dissipation process, see \cite{2006ApJ...642..902P}, but then one can certainly not use power-law wave spectra to fit the B/C data.

Meanwhile it is clearly desirable to consider alternative explanations for the peak in the energy-dependence of B/C, 
rather than blindly fitting the data with reacceleration; 
in particular convection gives a fairly natural mechanism, and there is evidence for a Galactic wind \cite{2012EAS....56...73E}.

\bibliographystyle{JHEP}

\bibliography{QERP}

\end{document}